\documentstyle[11pt]{article}

\input{epsfig.sty}

\def\boxit#1{
\vbox{\hrule height0.5pt\hbox{\vrule width0.5pt\kern10pt\vbox{
\kern10pt#1\kern10pt}
\kern10pt\vrule width0.5pt}\hrule height0.5pt}}

\def\bild#1\over#2{\mathrel{\mathop{\kern5pt #1}\limits_{#2}}}

\newcommand{\be}{\begin{equation}}
\newcommand{\ee}{\end{equation}}
\newcommand{\bc}{\begin{center}}
\newcommand{\ec}{\end{center}}
\newcommand{\ba}{\begin{array}}
\newcommand{\ea}{\end{array}}

\newcommand{\vx}{\vec{x}}

\newcommand{\br}{{\bf r}}

\newcommand{\bphi}{\bar{\phi}}
\newcommand{\bpi}{\bar{\pi}}
\newcommand{\bx}{ {\bf x}}
\newcommand{\bz}{ {\bf z}}
\newcommand{\bk}{{\bf k}}
\newcommand{\by}{ {\bf y}}

\newcommand{\lap}{$\lambda \phi^4$ }
\newcommand{\omek}{\sqrt{ {\bf k}^2 + \mu^2}}

\begin{document}

\begin{titlepage}
\title{\bf  Expanding non homogeneous configurations
of the $\lambda \phi^4$ model
 }
\author{  F\'abio L. Braghin\thanks{ email:braghin@if.usp.br} 
 \\
{\normalsize Instituto de F\'\i sica da Universidade de S\~ao Paulo} \\
{\normalsize C.P. 66.318,  C.E.P. 05315-970, S\~ao Paulo,      
Brasil }
}
\date{}
\maketitle
\begin{abstract}
A time dependent variational approach is considered to derive the
equations of movement for the $\lambda \phi^4$ model.
The temporal evolution of the model is performed numerically in 
the frame of the Gaussian approximation in a lattice of 
1+1 dimensions given non homogeneous initial conditions (like bubbles) 
for the classical and quantum parts of the field which expands. 
A schematic model for the initial conditions is presented 
considering the model at finite fermionic density.
The non zero fermionic density may lead
either to the restoration of the symmetry or to an even more 
asymmetric phase.
Both kinds of situations are considered as initial conditions 
and the eventual differences in early time dynamics are discussed.
In the early time evolution there is strong energy exchange
between the classical and quantum parts of the field as the
initial configuration expands.
The contribution of the quantum fluctuations
is discussed especially in the strong coupling constant limit.
The continuum limit is analyzed.
\end{abstract}



 IFUSP- /2001.

\end{titlepage}

\section{ INTRODUCTION }

There are many motivations for the study of 
time dependent non perturbative methods in Quantum Field Theory (QFT). 
Some examples are immediately found in systems with strong coupling 
constant, spontaneously symmetry breaking and which undergo phase
transitions.
Some of the most interesting cases are present in the 
relativistic heavy ion collisions currently being 
prepared and performed in BNL/RHIC 
which  probe hadronic matter at very high 
densities and temperatures. In these cases a region of
very high  energy  density starts expanding (and ``cooling'') 
just after the nuclei collision. 
These systems are usually described by hydro-dynamical 
models which are known to be quite reliable \cite{HYDRO}.
However, the understanding of these models in terms of microscopical
 descriptions
with the underlying QFT degrees of freedom are expected to be
derived and are currently being investigated by several groups. 
Many effects are expected to occur in those systems such as, for example,
dynamical  phase transitions whose existence should be present
in a reliable description.

Due to the extreme complexity of realistic theories, as QCD, 
one usually is lead to study effective models which respect the major 
properties of the fundamental theory and reproduce the main issues of it
in some range of a relevant variable (as energy).
In the present work, however, a still more simplified version
of the reality is considered in order to check qualitative effects.
The \lap model is often used as a test model, although
its scalar field may be considered as the relevant degree of freedom 
for inflationary models in Cosmology \cite{COSMO}. 
In condensed matter and statistical mechanics it is also
of interest \cite{CONDMATT}.
Furthermore
it can be identified to the
mesonic sector of the linear sigma $O(N)$ model in the large
$N$ limit or without pions.

In the last decade a quite large amount of work have been done 
in order to shed light in some of the subjects mentioned 
above as well as aspects related
to the dissipation, thermalization, decoherence, formation of disoriented
chiral condensates (DCC),
 phase transitions among others with non perturbative time dependent 
formalisms. Some examples are found in
\cite{EBOJAPI,COPISTA,COHAKLUMO,BOVE1,BOVEHO,FLB98a,BAHEPA},
\cite{SSV,BEGLERAM,AAAB,TSUE,BRANAV,AARTSSMIT}.
Some works have already been done concerning the dynamics of 
non homogeneous configuration in bosonic fields.
In particular, the Gaussian equations of movement were considered in 
\cite{BOVEHO,FLB98a,AARTSSMIT} to study properties of the large time
dynamics in certain cases.
Non homogeneous field theory was also studied in \cite{AARTSSMIT} 
to obtain information relevant for  
particle production and thermalization of non-equilibrium systems.

In the present work we  extend, complement and give a sounder 
picture for 
 the work done in \cite{BRANAV}.
The main aspects we study in the present work are the following.
Firstly a schematic model for the formation of locally 
non equilibrium initial conditions (which expands with time) is developed.
As already proposed in this last paper, we consider the
possibility that, due to some
particular condition,  the condensate amplitude
(as well as the physical mass of the scalars) is either suppressed 
or enhanced in a localized region of space. 
For this we consider the \lap model coupled to fermions at finite density
in a one dimensional space.
Depending on the kind of the coupling the model may experience
either symmetry restoration at high fermionic density or 
further symmetry breakdowns \cite{FLB2001b}.
We look for  dynamical consequences of the corresponding 
enhancement or suppression of the condensate at the tree
level and in the frame of the Gaussian approximation. 
Although there are enormous differences
between this simple (idealized) model (the \lap model)
and the realistic  ``fireballs'' from 
RHIC, we believe that the study of the 
influence of the present field theoretical degrees of 
freedom is of interest.
A partially similar idea to this one was discussed 
in reference \cite{DUMITRU} where the pressure due to 
a gas of pions, including a \lap term in a static description, 
was considered to  drive an expansion of a ``fireball''.

We have therefore the following picture.
Firstly we fix two  parameters of the model, which will allow us to 
make meaningful comparisons between the (time dependent)
tree and Gaussian levels. 
Secondly, it is assumed that the scalar field is locally
placed in a thermal bath and/or experiences an interaction
 with a finite fermion density in that small region.
These interactions -which change the ground state of the
model- are switched off at t=0 
yielding non equilibrium
initial conditions for the scalar field which are evolved in time.
The temporal evolution is performed within the 
tree level and Gaussian approach equations, producing
the expansion of regions (bubbles) endowed with high energy densities.

The work is organized as follows. In section 2 a
time dependent variational method for pure states systems 
is described in the 
Schr\"odinger picture and the equations of movement are derived
with a Gaussian trial wave functional.
Some considerations for the  static and thermal case are done.
The small amplitude motion case is investigated for 
homogeneous configurations in order to provide some 
useful results for analyzing the relevance of 
the one dimensional lattice simulations for a more
realistic situation.
In section 3 the numerical method using a pure states
generalized density matrix - Time Dependent Hartree Bobogliubov (TDHB), 
as developed in \cite{FLB98a}-
is extended for the asymmetric case ($\bphi\neq0$) 
in a discretized space.
It allows for the investigation of the temporal evolution of 
a localized region of the space (lattice) where a high energy 
density occurs.
Next, in section 4, 
by coupling the scalar field to a finite fermionic density system
we are able to construct a model for the initial conditions at
finite density.
The possibility of symmetry restoration and further asymmetric phase(s)
is discussed and the temporal evolution of corresponding 
configurations will be analysed.
Still in the section 4 we provide an alternative way of 
fixing the parameters of the model in order to perform
plausible comparisons between classical and quantum field systems. 
The numerical results for the early time evolution are 
presented in section 5 for different 
initial conditions and values of the free parameters of the model.
All the numerical examples shown in this article conserve total energy.
Special attention is given to the strong coupling limit and 
one example of the temporal evolution of
a deviation from a Kink solution is exhibited.
The continuum limit is discussed.
The results are summarized in the last section.

\section{ GAUSSIAN APPROXIMATION }

The time dependent variational approximation at the Gaussian level 
 has been studied for several years 
\cite{COPISTA,BOVE1,FLB98a,COHAKLUMOPAAND,SAPI}. 
It provides a systematic method to study the temporal evolution 
of a quantum field theoretical model given an initial condition
by means of the equations of movement. 
In spite of recent achievements for considering approximations 
beyond the Gaussian (for example in \cite{SSV,BERCOX})
we want to address some
unexplored features in the time dependent Gaussian approach.

Let us consider the time-dependent Dirac's variational 
principle with a trial Gaussian 
wave functional \cite{JACKER,COOPMOT}.
Firstly the average value of an action $I$ is calculated with
a given trial wave functional $|\Psi>$:
\be \label{ACTION} \ba{ll}
 I = \int d t <\Psi| \left( i \partial_t - H \right) |\Psi>,
\ea
\ee
where $H$ is the  Hamiltonian.

In the Gaussian approximation at zero temperature the wavefunctional
$\Psi$ is parametrized by:
\be \label{4} \ba{ll}
\displaystyle{
\Psi\left[\phi(\bx )\right] = N  exp \left\{ -\frac{1}{4} \int
d \bx d \by \delta\phi(\bx )\left(G^{-1}(\bx ,\by) + i\Sigma(\bx,\by)
\right) \delta \phi(\by)
+ i\int d\vx \bar{\pi} (\bx)\delta \phi (\bx) \right\} . }
\ea
\ee
Where $\delta\phi(\bx, t) = \phi(\bx)-\bar\phi(\bx,t)$;
 the normalization is  $N$, the variational parameters are   
the condensate $ \bar \phi (\bx , t) = < \Psi | \phi | \Psi > $ and its 
conjugated variable $ \bar{\pi}(\bx, t) = < \Psi | \pi | \Psi > $;
quantum fluctuations represented by the width  of the Gaussian  
$ G(\bx,\by,t) = <\Psi |\phi(\bx) \phi(\by)  | \Psi> $ and its conjugated
variable $ \Sigma(\bx,\by,t)$.

The Lagrangian for a scalar field   $\phi$ with bare mass   $m_0^2$
and quartic coupling constant  $\lambda$ is given by:
\be \label{1} \ba{ll}
\displaystyle{
{\cal L}(\bx) = \frac{1}{2}\left\{
\partial_{\mu} \phi(\bx) \partial^{\mu} \phi(\bx) - m_0^2 \phi(\bx)^2 -
\frac{b}{12} \phi(\bx)^4 \right\}.  }
\ea
\ee
From this expression the corresponding Hamiltonian $H$ is obtained.
The action of the field operator $\phi$ and its conjugated momentum
$\pi$ in functional Schroedinger representation 
over a wave functional $\Psi \left[ \phi (\bx) \right] = 
< \phi (\bx) | \Psi \left[\phi \right]>$ is respectively:
\be \label{2a} \ba{ll}
\displaystyle{ \hat{\phi} |\Psi \left[ \phi(\bx) \right] >
= \phi(\bx) |\Psi \left[\phi(\bx) \right] >,}\\
\displaystyle{ \hat{\pi} |\Psi \left[\phi (\bx)\right] >  =
- i \delta / \delta \phi (\bx) |\Psi \left[\phi(\bx) \right] > . }
\ea
\ee
The average value of the Hamiltonian, which will be explored in the 
numerical simulations, is given in terms of the variational
parameters with the functional integrations:
\be \label{11a} \ba{ll}
{\cal H} & = \frac{1}{2} \left[ \frac{1}{4} G^{-1}(\bx,\bx) + 
              4 \Sigma G\Sigma (\bx,\bx) +\bpi^2(\bx) 
             - \Delta G(\bx,\bx) + m^2_0 G(\bx,\bx) + \frac{\lambda}{4} 
              G^2(\bx,\bx) + \right. \\
   & \left. + m^2_0 \bphi^2 + (\nabla \bphi(\bx))^2 
     + \frac{\lambda}{12}\bphi^4(\bx)
      + \frac{\lambda}{2} \bphi^2(\bx) G(\bx,\bx) \right].
\ea
\ee
In the Schr\"odinger picture the wave 
functional evolves like the Schroedinger 
equation:
\be \label{3}
\displaystyle{ i \frac{ \partial}{\partial t} \Psi \left[\phi (\bx)\right] =
H \Psi \left[\phi (\bx)\right]. }
\ee
This equation is equivalent to the temporal evolution of the variational
parameters given initial conditions.

The variations of the $I$ with relation to the variational 
parameters (of the Gaussian wave functional) produce the equations
of movement.
\be \label{7gau} \ba{ll}
\displaystyle{ \frac{\delta I}{\delta \Sigma(\bx,\by,t) }
 \rightarrow  }&
\displaystyle{
\partial_t G(\bx,\by,t) = 2  \left( G(\bx,\bz,t)\Sigma(\bz,\by,t) +
 \Sigma(\bx,\bz,t)G(\bz,\by,t) \right)  }  \\
\displaystyle{ \frac{\delta I}{\delta G(\bx,\by,t)} 
 \rightarrow } &
\displaystyle{ 
\partial_t \Sigma(\bx,\by,t) =  \left(
2 \Sigma(\bx,\bz,t)\Sigma(\bz,\by,t) -
\frac{1}{8}G^{-2}(\bx,\by,t) \right)  + }\\  &
 \displaystyle{ + \left( \frac{\Gamma(\bx,\by,t)}{2} +
\frac{ \lambda}{2} \bar \phi(\bx,t)^2 \right)  } \\
\displaystyle{
\frac{\delta I}{\delta \bar{\pi}(\bx,t)} 
 \rightarrow } &
 \displaystyle{
\partial_t \bar \phi(\bx,t) = - \bar{\pi}(\bx,t)  }  \\
\displaystyle{
\frac{\delta I}{\delta \bar{\phi}(\bx,t) }  \rightarrow } & 
\displaystyle{
\partial_t \bar{\pi}(\bx,t) = \Gamma(\bx,\by,t)\bar \phi(\by,t) +
\frac{\lambda}{6}\bar \phi^2(\bx,t) }
\ea
\ee
where: $\Gamma (\bx,\by,t) = -\Delta  + \left( m_0^2 + 
\frac{ \lambda}{2}G(\bx,\bx,t)\right)\delta (\bx-\by)$.
In the x-coordinate space the Gaussian width is written as:
\be  \label{10}
G({\bx},{\by})=<{\bx}|\frac{1}{\sqrt{-\Delta + \mu^{2}}}|{\by}>.
\ee

As it is well known the static limit of this approximation 
produces the ``Cactus'' diagrams   resummation 
\cite{BARGAN}. 
In section 3, an alternative way of writing these equations
will be discussed and evolved in time in a lattice for 
a class of non homogeneous initial conditions. 
It corresponds to the time dependent Hartree Bogoliubov
approximation \cite{FLB98a}.

By performing a Fourier transformation it is possible to 
 eliminate the variables $\bar{\pi}$ and $\Sigma$. 
The equations in the asymmetric phase become:
\be \label{9} \ba{ll}
\displaystyle{ \ddot{G}_{k k'}  -
\frac{ \dot{G}^2_{k k'} }{2 } G_{k k'}^{-1} -
\frac{ 1 }{2} G_{k k'}^{-1} + 2 \left( k^2  + m_0^2  +
\frac{b}{2}G(x,x) + \frac{b}{2} \bar{\phi}
\right) G_{k k'} = 0   } \\
\displaystyle{ \ddot{\bar{\phi}}_k + \left( k^2
 + m_0^2  + \frac{b}{6} \bar{\phi}^2 + \frac{b}{2}G(x,x) \right)
\bar{\phi}_k = 0               }
\ea
\ee
with  $ G_{\bk \bk'} = <\bk|G(\vx,\vx)|\bk +{\bf q}>$. 
These equations
were generalized for the out of equilibrium (non zero temperature) 
using different methods in 
\cite{EBOJAPI,BAHEPA,BEGLERAM,COHAKLUMOPAAND,BERCOX}
and they were studied extensively mainly 
by means of numerical calculations.
The choice of initial conditions
is entirely subordinate to the approximation, in the sense that
were it not Gaussian one might have to consider three conditions instead of
two \cite{COOPMOT}. 
The analysis of these equations in  show that initial conditions 
(for homogeneous $G$ and $\bphi$ ) are  crucial for the time interval 
in which the system evolves towards the minimum and beyond as well as  
for the speed of the field evolution \cite{BOVE1,BOVEHO}. 
Some other effects have been addressed as, for example, dissipation via 
particle production, the Landau damping,  
collisional relaxation at zero temperature 
as well as the relevance of the initial conditions for the
dynamics was also investigated in some cases in the early and late 
time evolution
\cite{COHAKLUMO,BOVE1,FLB98a,BRANAV,PIZA,AARTS,BEPASA,DEMA}.
In particular, it was found that the Hartree approximation is 
well suited for the study of early-time dynamics.

\subsection{VACUUM AND RENORMALIZATION}

The state of minimum energy is found from the equations of movement 
in the static case $\dot{G} = {\Sigma} = \dot{\bphi} = 0$. 

The equations of motion reduce to the GAP equations which minimize the 
effective potential.
They can then be written as:
\be  \label{9a} \ba{ll}
&\displaystyle{ \bphi \left( \frac{\lambda}{6}\bphi^2 + m_0^2 + 
\frac{\lambda}{2} {G} (\mu^2) \right) = 0
,}\\
&\displaystyle{ \mu^2 = m^2_0 + 
\frac{\lambda}{2} \left( \bphi^2 + {G}(\mu^2) \right) ,}
\ea
\ee
These equations provide the two phase \lap model: a symmetric phase
(where there is only a zero condensate $\bphi =0$) 
and the asymmetric phase where the 
condensate in non zero in the vacuum.
From these equations, in the asymmetric phase, we find that:  
\be \label{212} \ba{ll}
\displaystyle{ \bphi^2= \frac{3\mu^2}{\lambda},}
\ea
\ee
In spite of written in terms of the bare coupling constant 
this value  can be compared to the 
tree level value $\bar{\phi}=\sqrt{-\frac{6m_0^2}{\lambda}}$.
In the former case (expression (\ref{212})) 
the bare coupling constant is fixed by the 
value of the renormalized coupling, as discussed below. 

For a thermal environment, in d dimensions, we can perform a similar
calculation to the above one with a density matrix with mixed states.
It yields thermal fluctuations corrections to the two point Green's function
$G(\bx,\bx)$ which is substituted by: 
\be \label{intdiv} \ba{ll}
\displaystyle{ \tilde{H}(\mu^2) 
 = \int \frac{d {\bf k}}{(2\pi)^d} <\phi^2>_k = 
\int \frac{d {\bf k}}{(2\pi)^d} \frac{1+ coth\left( \frac{\beta \omek}{2} 
\right) }{\omek} 
= \int \frac{d {\bf k}}{(2\pi)^d} \frac{1}{\omek} 
\left( \frac{1}{2} + f(\bk) \right). 
}
\ea
\ee
In the above expression $f(\bk)$ is the Bose-Einstein occupation number.
It is well known that these thermal fluctuations for the asymmetric 
phase may restore the symmetry: the vacuum solution of the condensate
$\bphi_0$ tends to become zero as temperature increases.
At very high temperatures ($T>>\mu$) the integral (\ref{intdiv}) 
in one spatial dimension can 
be expanded and the second expression of (\ref{9a}) can be 
written in one spatial dimension as:
\be \label{expansion} \ba{ll}
\displaystyle{ \mu^2_T = \mu^2 - \frac{\lambda}{4}
\left( \frac{1}{2\beta \mu} + \frac{1}{2\pi} (ln(\mu/(4\pi T) 
+ \gamma ) \right) + {\cal O}(T^2), }
\ea
\ee
where $\gamma=0.57721$ and $\mu^2$ is the mass at zero temperature.
Therefore 
the physical mass is reduced at high temperatures, 
$\mu^2_T = \mu^2(T) < \mu^2$. 
This effect is also present in the condensate (by means of 
expression (\ref{9a})) making possible 
the eventual restoration 
of the symmetry breaking (at least at this level of 
approximation) \cite{RSSB}.

The renormalization of the time dependent approach (in continuum
spaces) can be done
in the same way at zero and finite temperatures as well as 
out-of-thermodynamical equilibrium \cite{BAHEPA,COHAKLUMOPAAND,SAPI}. 
No additional ultraviolet divergence is found 
due to temporal evolution of the system or its departure from 
zero temperature or equilibrium. 
Renormalization can therefore be performed by absorbing the 
divergences in the 
physical mass (1 dimensional system) in the vacuum. 
By re-writing the GAP equations in terms of the renormalized quantities
($m^2_R, \lambda_R$)
we can find the relationship between the bare and physical quantities. 
They can be written in one dimension as:
\be \label{331d} \ba{ll}
\displaystyle{ m_R^2 (1d) = m_0^2 + \frac{\lambda}{8 \pi} ln\left( 
\frac{2\Lambda}{\mu} \right)
,} \\ 
\displaystyle{ \lambda_R (1d) = \lambda \left(  
\frac{1- \frac{\lambda}{8\pi m_R^2} }{1+ \frac{\lambda}{16\pi m^2_R}  }
\right)
.}
\ea
\ee
We note that the coupling constant acquires only a finite correction. 

In a discretized space there is a natural regulator 
 (the lattice spacing for which:
$\Lambda = 1/\Delta x$). 
The integrals (\ref{intdiv}) become a summation and converge. 
As the limit to the continuum is 
taken the integrals tend to diverge and a redefinition of 
the bare quantities is needed. 
This will be discussed in section 5.

\subsection{ SMALL DEVIATIONS AROUND THE MINIMUM }

Before exhibiting numerical solutions of the equation of movement 
in a lattice let us perform an exercise which is useful for the 
understanding of results. 
For a certain class of initial conditions it is possible 
to find analytical solutions for the equations of movement.
Firstly because we can see the relevance of the temporal 
evolution of a given initial condition irrespectively to the
regularization method for the local divergences.
We can also check that the study of the one dimensional 
model can provide information about the three dimensional case. 

We therefore consider the symmetric phase of the model ($\bphi = 0$) 
assuming a particular kind of homogeneous initial  conditions 
(exactly the same calculation can be
done for non-homogeneous asymmetric potential 
cases, this would not change the 
conclusions).
Let us assume that, for some reason external to the model, 
the system experiences a small deviation from the state of minimum.
As discussed before, 
this can happen because of an external thermal bath 
and it can be parametrized by a change in the physical mass, e.g.
$m(t=0)=.9\mu$.
This  drives quantum fluctuations away from the ground state and 
generates dynamical evolution. 
In this case it is possible to extract exact analytical 
solutions for the equation of movement. 
For this kind of initial conditions one 
can linearize the Gaussian equations of motion 
using the following  prescription for the solution:
$$G(m^2,t) = G_0(\mu^2) + \delta G(t),$$ 
where $G_0(\mu^2)$ 
is the value of the fluctuations which obey the GAP equation 
 and $\delta G(t)$ the value of the 
small deviation which evolves in time.

We have presented analogue solutions for the free case in
\cite{FLB98a} choosing a plane wave prescription for the deviation
$\delta G(t)$ since an infinite system is considered.
We have found the following solution for the 
deviation $\delta G(t)$ in d-dimensions:
\be \ba{ll}
\displaystyle{ \delta G(t) = \frac{\mu^2-m^2}{2} 
\frac{ I_{3c}}{1-\lambda I_3 - \frac{\lambda}{2}I_{3c} } , }
\ea
\ee
where:
\be \ba{ll}
\displaystyle{ I_3 = \int \frac{d^d \bk}{(2\pi)^d} 
\frac{1}{(2\sqrt{\bk^2+\mu^2})^3} \;\;\;\;\;\;\;\;
I_{3c}  =  \int \frac{d^d \bk}{(2\pi)^d} 
\frac{ cos(2\sqrt{\bk^2 +\mu^2} t) }{(2\sqrt{\bk^2+\mu^2})^3 } .}
\ea
\ee
The integral $I_3$ is time independent and contains a log divergence in
3 dimensional space.
It can be  absorbed in the renormalization of the mass and 
coupling constant.
The integral $I_{3c}$ also has a divergence at $t=0$ in three 
spatial dimensions which also can be regularized.  
This problem 
was also addressed in \cite{FLB98a,BAABOVE}.
In particular, the integrals can  be written in terms of 
generalized hypergeometric functions.

In an one dimensional space, on the other hand, there is no 
(ultraviolet-UV) divergences for this case of temporal evolution. 
Given a finite initial condition (which is the deviation
from the vacuum value, proportional to  $m^2 - \mu^2$) 
the temporal evolution is finite. 
This is very significant for the numerical solutions
in a lattice: 
the temporal evolution of the system is independent of 
the regulator.
This happens due to the fact that the static sector was
separated from the dynamic evolution and the
divergences only contributed for the redefinition of the former.
In other words, the regularization and renormalization do not
interfere in the temporal evolution. 
This is checked in section (5) with numerical calculations
 in a 1-dim lattice.

Other interesting feature can be pointed out from these solutions
by rewriting them.
In an 1 dimensional space 
the time dependent integral can be written as:
\be \label{i3c1} \ba{ll} 
I_{3c}^{(1)} = \frac{(\mu^2 -m^2)}{8\pi} {\bf \int^{\infty}_{\mu} d y 
\frac{cos(2yt)}{y^2 \sqrt{y^2- \mu^2} } }.
\ea
\ee
Whereas in three spatial dimensions the same 
time dependent integral can be written as:
\be \label{i3c3} \ba{ll}
I_{3c}^{(3)} = \frac{(\mu^2 -m^2)}{8\pi} 
\left( \mu^2 {\bf \int^{\infty}_{\mu} d y 
\frac{cos(2yt)}{y^2 \sqrt{y^2- \mu^2} } } + \frac{\pi N_0(2\mu t)}{2}  
\right),
\ea
\ee
The main difference between these two expressions 
in the fact that in three dimensions there are
two additional divergences, one
of them occurs only at $t=0$.
This also suggests  
that the dynamics in one spatial dimension is not 
completely different from
the dynamics in three dimensions for the cases under study.

\section{ NUMERICAL METHOD: Hartree Bogoliubov}


This section is an extension of the work developed in 
\cite{FLB98a} for  the time dependent Hartree Bogoliubov approximation
in the presence of a (time and space dependent) condensate.
The  generalized Hartree-Bogoliubov density matrix can be defined, 
in a discretized space, as \cite{BLARIP}:
\begin{eqnarray} \label{Rij}
     R_{i,j} =   \left(
        \begin{array}{ll}
          \rho_{i,j}  &  \kappa_{i,j}   \\
       -\kappa^{\ast}_{i,j}  & -\rho^{\ast}_{i,j}
         \end{array}
        \right),
\end{eqnarray}
where the average density matrices are written in terms of the 
average of creation and annihilation operators:
$\rho_{i,j} = \frac{1}{2} <a_i a^{\dag}_j
+ a^{\dag}_i a_j > $ which is hermitian and 
$ \kappa_{i,j}= -<a_i a_j > $ is symmetric.

For the calculation of this matrix,  
the creation and annihilation operators are written, 
 in a d spatial dimensions lattice, in terms of the 
field and its conjugate variable:
\begin{eqnarray}
a(j) & = & \frac{1}{\sqrt{2}}\left\{ \phi(j)(\alpha)^{\frac{d-1}{2}} +
   i \pi(j)(\beta)^{\frac{1+d}{2}} \right\} \\
a^{\dag}(j) & = & \frac{1}{\sqrt{2}}\left\{ \phi(j)
(\alpha)^{\frac{d-1}{2}} -
   i \pi(j)(\beta)^{\frac{1+d}{2}} \right\},
\end{eqnarray}
Where $\alpha$ and $\beta$ are  (dimensional) normalization factors. 
A suitable choice for them is the lattice spacing.

The averaged values of the field variables ($\phi,\pi$)
previously shown can be written in terms of the matrix elements of the 
above generalized  matrix $R$. 
Using the notation of expressions (\ref{7gau},\ref{Rij}), 
we obtain at zero temperature:
\be \label{63} \ba{ll}
\displaystyle{ <\phi(i) \phi(j) > =
G(i,j) + 1 \;\; \bphi_i^2 = 
 \frac{1}{(\Delta x)^{d-1} }
\Re e ( \rho(i,j) + \kappa(i,j) )   }\\
\displaystyle{ <\pi(i) \pi(j) > =
F(i,j) + 1 \;\; \pi_i^2 = 
 \frac{1}{(\Delta x)^{d+1} }
\Re e ( \rho(i,j) - \kappa(i,j) ), }\\
\displaystyle{ < \phi(i) \pi(j) > =
2 G(i,k)\Sigma(k,j) + \bphi_i \bpi_i
= -2 Im (\rho(i,j) -\kappa(i,j) ),}\\
\displaystyle{ < \pi(i) \phi(j) > = 
2 \Sigma(i,k)G(k,j) + \bpi_i \bphi_i
=  2 Im (\rho(i,j) +\kappa(i,j) ),}\\
\ea
\ee
where $ F(i,j) = G(i,j)^{-1}/4 + 4 \Sigma(i,k)  G(k,l) \Sigma(l,i)$.
Therefore we can already note that there are two different ways of 
expressing the state of the system: by means of the generalized 
density matrix elements OR with the variational 
parameters discussed before. 
{\it They are completely equivalent descriptions for the prescriptions
used in this work}.

Next, we discuss the dynamics of the generalized density matrix. 
The temporal evolution is governed by the Hartree- Bogoliubov energy, 
which can be parametrized in the following form:
\begin{eqnarray} \label{64a}
  \frac{1}{2} H_{ij} \equiv \frac{\delta E}{\delta R_{ji} }=
          \left(
                     \begin{array}{cc}
                    W_{i,j}  & D_{i,j} \\
                   -D_{i,j} & -W_{i,j}
                      \end{array}
                  \right).
\end{eqnarray}
It is worth to notice that this approximation, and consequently the 
dynamics under study, is invariant under an unitary transformation.
Therefore it is possible to consider the generalized density matrix and 
Hartree Bogoliubov ($H_{i,j}$) energy in another form, namely:
\begin{eqnarray} \label{64}
\tilde{R} = \frac{1}{\sqrt{2}} \tau R 
             \frac{1}{\sqrt{2}} \tau^T,\;\;\;\;\; 
\tilde{H} = \frac{1}{\sqrt{2}} \tau H
             \frac{1}{\sqrt{2}} \tau^T.
\end{eqnarray}
where the superscript 
$^T$ means the transposed matrix and $\tau$ is given by:
\begin{eqnarray}
  \tau =
          \left(
                     \begin{array}{cc}
                    1  & -1 \\
                   1  &  1
                      \end{array}
                  \right).
\end{eqnarray}
With these transformed matrices, one can check that a  Liouville-von 
Neumann type equation is necessarily satisfied
only in the symmetric phase
($\bphi=0$):
\be \label{64b} \ba{ll}
i \dot {\tilde{R}}_{i, j} 
= \left[ \tilde{H}_{i, k} , \tilde{R}_{k, j} \right].
\ea
\ee
This equation can be written in terms of the averaged quantities (\ref{63}).
By doing this exercise we have shown that they are equivalent
to equations (\ref{7gau}) in the zero temperature limit without 
the condensate.
When the classical field is taken into account, the
 equation (\ref{64b}) is still considered and 
$\bphi(x)$ acts  as an external
(dynamical) source
to (\ref{64b}). 
The time dependence of $\bphi (t)$ is determined by 
the two last equations of the set (\ref{7gau}). 
These last equations, on their turn, 
depend on the fluctuations 
which are evolved in (\ref{64b}). 
We are currently extending this method to finite temperature 
non equilibrium systems  \cite{FLB2001b}.
The numerical evolution of (\ref{64b}) is 
obtained by diagonalizing $R_{i,j}$ and 
evolving the eigenvectors from which the variational parameters 
(eg. $G_{i,j}$) can be calculated.

\section{PARAMETERS OF THE MODEL AND INITIAL CONDITIONS}

In this section we complete the schematic model for justifying the
 initial conditions used in the evolution of 
equations of movement. The finite temperature 
effects were already discussed in the frame of the Gaussian approach 
and now we couple the \lap model
to fermions at finite density. 
There are many possibilities for this coupling and we discuss
only a few cases which may lead either to the restoration of 
the spontaneously broken symmetry or to a
further asymmetric model at high densities.
We also provide some remarks which are helpful for the 
choice of the sets of parameters: physical mass $\mu^2$ and 
coupling constant $\lambda$ in the
lattice.

\subsection{\lap COUPLED TO FERMIONS AT FINITE DENSITY }

We can couple the \lap model to fermions yielding a 
mechanism for placing the system in a finite fermion density environment.
This procedure may produce another mechanism for the symmetry restoration.
In the dynamical picture we suppose that the 
coupling of fermions to the scalar field 
is switched off at $t=0$ as a first approximation to the problem.
In part, this can be justified in the case  one has
decreasing  scalars-fermions interaction amplitude
with relation to the self interaction of the scalar field. 
If we consider an expanding finite density environment (``fireball'')
the density is expected to decrease as the system expands.
As we will not be concerned with the
fermion dynamics we will only consider the interacting part of 
its Hamiltonian.

The following Hamiltonian density terms for spin half 
fermions $\psi$ coupled to the scalar field are considered:
\be \label{70INT} \ba{ll}
\displaystyle{ {\cal H}_I = g_a \phi (\bx)\bar{\psi}(\bx)
\psi(\bx) + g_b \phi^2(\bx)
\bar{\psi}(\bx) \psi(\bx) , }
\ea
\ee
where the coupling constants $g_a$ and $g_b$ are dimensionless
only in  3+1 and 2+1 dimensions respectively. 
Nevertheless we are allowed to consider them as 
effective couplings of an effective theory. 
Eventually one would need 
other couplings as one considers higher energy processes.

The above couplings lead to changes in the equations of the \lap
model.
In particular, we are interested in possible effects in the 
minimum energy state of the model. 
For this, we repeat the calculation
of the GAP equation which is obtained by the minimization
of the Hamiltonian density (expression \ref{11a}) in the frame of the 
Gaussian approximation. 
Considering that the wave functional
of the system now acquires a fermionic sector $|\Xi>$ (which
may be a Slater determinant and is a function of the 
chemical potential) due to 
the presence of fermions at finite density we write:
\be \label{70fun} \ba{ll}
\displaystyle{ | \Psi \left[ \phi \right] > \to 
| \Psi \left[ \phi \right] > \times | \Xi \left[ \psi \right] >
.}
\ea
\ee
With the  fermionic wave functional one calculates the averaged values
of the interacting Hamiltonian terms which enter in the 
effective potential of the scalar field and modify the 
GAP equations. 
These averaged terms are given in terms of a
fermionic density:
\be \label{70aver} \ba{ll}
\displaystyle{ <\Xi | \bar{\psi}(\br) \bar{\psi}(\br)
| \Xi > = \rho_f (\br)
.}
\ea
\ee
The density $\rho_f$ can be calculated by fixing the chemical potential
eventually at a given temperature. 
However it is not the aim of this work
to perform such a detailed dynamical self consistent microscopic description. 
Here we only want to provide a physical basis for the 
 initial conditions of the time dependent model.
By the minimization of the energy density 
with respect to the one and two point functions of the scalar field,
we obtain the 
GAP equations (at finite fermionic density) which can be written as:
\be \label{70gap} \ba{ll}
\displaystyle{ \mu^2 = m_0^2 + \frac{\lambda}{2} \left(
G + \frac{\bphi^2}{2} \right) + 2 g_b \rho_f 
,}\\
\displaystyle{ \left( \mu^2 - \lambda \frac{\bphi^2}{3} \right) \bphi
+ g_a \rho_f = 0
.}
\ea
\ee
Regularization of the ultra violet divergences and 
the subsequent renormalization of the parameters of the model
does not change qualitatively these equations.


We fix a coupling constant (a strong one:$\lambda \simeq 235 \mu^2$,
where $\mu$ is the physical mass) 
for an one dimensional system but
the conclusions do not change qualitatively for higher dimensions.
Solutions for these two expressions (for fixed couplings)
have been searched and the
main conclusions due to the introduction of the fermionic density
dependence through the couplings $g_a$ and $g_b$ are the following:

(i) Keeping $g_b=0$ and a quite small $g_a = + \mu/5$.
As can be seen in expression (\ref{70gap}) there will not exist the
symmetric solution $\bphi_0=0$ at non zero density. In fact
as the density increases the (real) condensate value will increase.
This leads to a still more asymmetric phase.

(ii) On the other hand for the same value of the other coupling 
$g_b = + \mu/5$ but $g_a=0$ the asymmetric phase will 
disappear when $\rho \simeq 14 \mu^2$ 
and there will be no more condensates. At nearly double densities
the physical mass also disappears.

(iii) If one assumes negative coupling 
$g_b < 0$ (and $g_a=0$). 
In this case the condensate has increasing values for higher 
densities  until 
a point where there is a complete disappearance of the 
asymmetric phase. This point coincides with a zero value for 
the mass from the GAP equation at $\rho \simeq 25 \mu^2$.

For finite temperature field theories, as O(N) or O(N)XO(N) models.
the spontaneously symmetry breakdown may be restored or not at high
temperatures \cite{NRSSB}.
We point out, however, the possibility of 
further symmetry breaking at finite densities.
A more complete analysis the finite density effects on 
scalar models will be shown elsewhere \cite{FLB2001b}.


We obtained, therefore, another mechanism for 
considering local variations of the condensate: $\bphi \neq \bphi_0$.
This allows for scenarios in which one obtains non homogeneous 
configurations for the scalar fields at finite fermionic densities
and temperatures fixing the initial conditions. 
The situation in which the condensate is suppressed
is usually more accepted.
We do not neglect the possibility of enhancing the condensate due
to some different phenomena when the field interact with matter.

\subsection{CHOICE OF PARAMETERS}

We consider static initial conditions, the initial 
``velocities'' of the classical and quantum parts of the 
field are taken to be zero in all examples shown below:
\be 
\Sigma \propto \dot{G} = 0 \;\;\;\;\;\;  
\pi \propto \dot{\bphi} = 0,
\ee
where the dot means time derivative.

The \lap model has two
 parameters which must be fixed: mass and coupling 
constant. 
We have chosen some values to place the system in
the scaling limit. 
With the lattice spacing $\Delta x = 0.1fm$ we had
considered $\Delta x << \xi < L$ where $\xi=1/\mu$ is the correlation
length and $L$ is the size of the lattice. In this region the 
universal properties of the lattice model can, in principle, 
be described by a continuum field theory.
As a rule, the physical mass 
was chosen to be $\mu = 100 MeV$ for the dynamical situations.

For the coupling constant, which 
has dimension $fm^{-2}$ in one dimensional space, 
different values were chosen:
from $\lambda = 1/12 \mu^2 \simeq 0.021 fm^{-2}$
to $\lambda=600 fm^{-2} \simeq 2350 \mu^2$.
For low dimensions the \lap model is super-renormalizable and the
coupling constant is large \cite{ZINNJUSTIN}.
The smaller value was already considered 
to be in the non perturbative regime \cite{SSV}.
We have found, however, that this is not the case for the examples
shown in the present work.
Indeed, we have found that the dynamics for couplings with values up to 
$\lambda \simeq 1 fm^2 \simeq 5/12 \mu^2$ 
are not substantially different from the tree level case in the
early times dynamics.
Numerical examples will be shown in section 5.

In order to perform consistent comparisons between classical
and quantum dynamics as well as among different initial
conditions it is important to fix the parameters of the model.
But instead of fixing physical mass and coupling,
it is also possible to fix other variables 
as the energy density or particle number.
For the sake of the argument, 
let us consider the simpler case of homogeneous 
solutions in the vacuum.
At the tree level ($G=0$) we obtain, in the vacuum:
\be \label{75a} \ba{ll}
\displaystyle{ {\cal H}_{vac} = -\frac{3 m_0^4}{2 \lambda}.
}
\ea
\ee
Therefore we can fix, for example, the mass and 
the  energy density 
of the vacuum and calculate  the corresponding coupling constant.
This can be useful for the study of the influence of the
quantum effects because the inclusion of fluctuations 
(perturbatively or not) change the ground state which is 
defined by the mass and coupling constant \cite{FLB2001b}.

Let us consider a regularized energy density 
${\cal H}_{G,vac}$ at the Gaussian level which 
can be particularly well suited for the lattice calculations.
Fixing the physical mass we obtain $G$. Writing the total energy 
density with expressions (\ref{11a}) and (\ref{212})
 we obtain a second degree polynomial 
expression for the coupling constant.
In terms of the bare (regularized) quantities the
corresponding solutions for the coupling constant as a function
of the (regularized) energy density 
are given by:
\be \label{76} \ba{ll}
\displaystyle{ \lambda = \frac{ -\delta \pm 
\sqrt{\delta^2 - 4\mu^4 G^2}
}{ 2 G^2 },
}
\ea
\ee
where $\delta = 8 {\cal H}_{vac} - G^{-1} + 2 \mu^2 G$.
The renormalized version of ${\cal H}$ (as derived, for 
example, in \cite{STEVENSON}) can also be used for this 
calculation.
In one dimension
there may have negative and complex solutions which
do not seem to correspond to meaningful stable minima of 
the effective potential \cite{FLB2001b}.
This calculation could be meaningful even in 3+1 dimensions
to the extent that the
\lap model can be considered as effective
for which the cutoff can be fixed at some
high energy scale.
In this case the bare and renormalized quantities 
can be related by expressions  (\ref{331d}) in one spatial dimension.
This is equivalent to placing the system in a lattice, which provides
us with a natural cutoff ($1/\Delta x$), and to perform all calculations
on it. 


\subsection{THE PICTURE}

We have therefore the following picture.
Firstly we fix two parameters of the model, which will allow us to 
make meaningful comparisons between the (time dependent)
tree and Gaussian levels. 
Secondly, it is assumed that the scalar field locally
 experiences an interaction
 with a finite fermion density in a small region or 
has some contact with a thermal bath.
These interactions -which change the ground state of the
model- are switched off at t=0 
yielding non homogeneous (and non equilibrium)
initial conditions for the scalar field which are evolved in time.
The temporal evolution is performed within the 
tree level and Gaussian approach equations, producing
the expansion of regions (bubbles) endowed with high energy densities.

\subsection{ INITIAL CONDITIONS: FINITE DENSITY AND TEMPERATURE}

Firstly we suppose that the temperature varies continuously from the center
of the lattice, where there is a high energy region, 
to the (zero temperature) vacuum. 
This constitutes an out of
equilibrium situation which 
 can be implemented in a lattice by the following configuration:
\be \label{IC1}
m^2(x,t=0) = \mu^2 tanh^2 \left(\frac{x-L/2}{A}\right).
\ee
The bubble of high energy density is centered at $x=L/2$ ($L$ being
the size of the lattice) and it has
size given by $A$ (which will be taken to be $0.5 fm$). 
At the center of this bubble $m^2 = 0 MeV$ which implies very high 
temperatures.
As the temperature also modifies the order parameter of the
model ($\bphi$) it may eventually yield a symmetry restoration.

We will consider
another kind of configuration given by:
\be \label{IC2}
\bphi (x,t=0) = \bphi_0 tanh^2 \left( \frac{x-L/2}{A} \right).
\ee
In the central region of the lattice
the condensate is suppressed. 
This can happen due to the interaction
with a finite density medium.
An unifying picture can be associated
to the above configurations. In  high energy collisions
there is a large amount of energy deposited 
in a small region, eventually making the
system to have its parameters modified, such as particle masses
and condensate. 
This energy excess 
is  expected to propagate (expand) in the space time.

An enhancement of the condensate due to the interaction with
matter can also be assumed (it could correspond to further 
symmetry breaking at high densities):
\be \label{IC3}
\bphi (x,t=0) = \bphi_0 \left[ 1 + \alpha 
sech^2 \left( \frac{x-L/2}{A} \right) \right],
\ee
Where $\alpha$ is a real positive number which measures the 
amount of energy excess deposited in the central region of the 
lattice.

Finally, another kind of initial conditions will be considered. 
It corresponds to 
a deviation from the KINK classical solution to the equations of
movement.
The static KINK solution of the classical \lap model is given by
\cite{MONTMUNST}:
\be \label{IC4} \ba{ll}
\displaystyle{ \bphi (x,t=0) = \bphi_0 tanh \left( \frac{x-L/2}{B}
\right),
}
\ea
\ee
where $B=\sqrt{-2/m_0^2}$. 
In this case 
$\bphi(r=0)= -\bphi_0$ and $\bphi (r=10) = \bphi_0$, where
$r=0$ and $r=10$ are the borders of the lattice. 
This configuration can be seen as a wall which separates
regions with different vacua where $\bphi = \pm \bphi_0$.
It is stable at the tree level and corresponds to a 
non trivial topology case.
By considering anti-periodic boundary
conditions we have chosen a deviation with relation to the 
kink solution:
\be \label{IC5} \ba{ll}
\displaystyle{ \bphi (x,t=0) = \bphi_0 tanh \left( \frac{x-L/2}{B}
\right).
}
\ea
\ee
Where $B$ was considered to be $B \simeq 1/(4 \mu)$. 
There is an energy excess in the central region of the lattice,
over the Kink.

\section{NUMERICAL RESULTS}

In Figure 1 some of the  initial conditions for the 
field ($G(r)$ and $\bphi(r)$) are shown. 
In thick solid line it is shown
$G=G(m^2(r),t=0)$  corresponding to a bubble of high energy 
density which comes out from
a zero mass region inside the vacuum (for $\mu=150 MeV$).
The physical mass is given by expression (\ref{IC1}):
$$ m^2(t=0) = \mu^2 tanh^2 \left(\frac{x-L/2}{0.5}\right); \;\;\;\;\;
\bphi (t=0) = \bphi_0.$$
In this case, it is therefore
assumed that the deviation of the mass 
in the center of the lattice
but the condensate is kept constant. 
The other two curves (thin solid and dotted lines)  
correspond to 
non trivial initial conditions for the condensate (\ref{IC2},\ref{IC3}) 
and  will be discussed later.

The first initial condition (\ref{IC1}) 
is evolved for two different values of the  coupling constant:
$\lambda = \mu^2/12 \simeq 0.02 fm^{-2}$ and $600 fm^{-2}$, for a physical mass
$\mu = 100 MeV$. 
The resulting evolution of  the energy density configuration 
for low times 
are shown respectively in Figures 2  and 3.
In figure 2 the energy density configuration is exhibited at different
time steps:
$t= 0.1fm, 1.0fm, 2.0fm$ and  $3.5fm$ as a function of 
the lattice points. 
 The initial (higher) energy bubble expands
(by "waves") towards the extremities of the lattice distributing 
the initial "potential energy" over the lattice and among 
classical and quantum degrees of freedom.
For a much stronger coupling constant, figure 3, 
the  energy density configurations are shown
and expand in a more ``concentrated way'' 
than in the weaker coupling case.
Almost no difference can be seen 
in what concerns the expansion velocities.

The following initial condition is considered for the next 
cases (expression (\ref{IC2})):
$$ \bphi (t=0) = \bphi_0 tanh^2 \left( \frac{x-x_0}{0.5} \right),
\;\;\;\;\; \dot{\bphi}=0 .$$
In this case the condensate is suppressed in the central region
of the discretized space-time.
In figure 4 the evolution of the corresponding energy density is shown 
for a quite weak coupling constant 
$\lambda = 0.6 fm^{-2} \simeq 2.36 \mu^2$
 without quantum fluctuations, i.e., by
means of the classical equation of motion -third and fourth equations 
of  (\ref{7gau}).
Two-peak waves expand from the center of the lattice to the borders.
The initial energy density has a two-peak structure which
happens due to the fact that the gradient term in the 
Hamiltonian density  has the largest values (see expression
(\ref{11a})).
This figure can be compared to the evolution of 
the complete system including quantum fluctuations which is shown
in figure 5. The two point function $G$ is considered to be in the
vacuum value at $t=0$. Due to the weakness of the coupling constant
(the same as the one used in figure 4) there is no
visible difference between both figures. 
The main difference, which is 
quite big but not relevant to the dynamics, is the overall normalization
of the energy density. The inclusion of quantum fluctuations
introduces a zero point energy whose effect on the dynamics is not
relevant, as the comparison between figures 4 and 5 shows.
The above value of the coupling constant still is in the 
(dynamical) perturbative regime for this early time analysis.
Comparison of this weak coupling constant case 
($\lambda \simeq 1/12 \mu^2$) to other works \cite{SSV}
suggests that the non perturbative dynamics may be present for some
kind of configurations but not for others in the early time analysis.

We consider now another type of initial condition. The 
value of the condensate is enhanced in the central region of 
the lattice by means of expression (\ref{IC3}):
$$
\bphi (t=0) = \bphi_0 \left[ 1 + 
\alpha sech^2 \left( \frac{x-L/2}{A} \right) \right],$$
where $\alpha$ is a real number which will be assumed to be in the
range $0 < \alpha < 0.5$. 
It measures the energy of the bubble
which is placed in the center of the lattice. 
It still  has size roughly
given by $A = 0.5 fm$. 
Since the previous examples have showed that $\lambda = 0.6 fm^{-2}
\simeq 2.36 \mu^2$ is 
not strong enough as to yield non perturbative quantum effects
we consider a stronger value
($\lambda = 60 fm^{-2} \simeq 236 \mu^2$) for the next figures.
This configuration 
could correspond either to a case
the model is place in a region with finite fermion density
with the particular kind of coupling discussed in the previous section
or to a situation where external fields
would ``merge'' in a condensate in that small region increasing
its value and the respective energy density  \cite{BRANAV}.

In figure 6 the classical energy density is shown for an initial
condition given above with $\alpha = 0.2$, which represents a 
not very energetic bubble. The energy density evolves nearly as the 
same way as the case of figure 4 (initial condition defined by
a suppression of the condensate instead of enhancement). 
There is a small 
 difference which is related to the relative amount of 
energy in each peak of the expanding waves at larger times (when
$t\simeq 3.5 fm$). In the present case (enhancement of the condensate)
the second peak is slightly higher with relation to the configuration 
 where the condensate is suppressed (figure 4). 
By switching on quantum fluctuations we obtain
the temporal evolution shown in figure 7.
The first issue we can note is that between time steps $0.1fm$
and $1.0fm$ there is an ``amplification'' of the locally concentrated
energy density  leading regions in which its value is smaller than
the vacuum energy density. Theses regions expand.
 These are local
effects since the total energy is always conserved and given by the
vacuum energy plus the initial bubble energy excess (positive).
Furthermore, these regions tend to disappear for larger times.
Another important effect is that the most part of the expanding 
energy is now concentrated in the ``advanced'' peaks, which arrive 
first in the extremes of the lattice. This corresponds to an 
acceleration of the expansion \cite{BRANAV}.
For the preceding cases the energy density ``bumps'' took nearly
$\Delta t \simeq 5$ fm to arrive at the borders of the lattice.
In the present case (figure 7), 
 this time is reduced to nearly $\Delta t \simeq 3.8$ fm.
As noted in this last reference the energy density becomes
smoother when considering the quantum fluctuations but this
happens only in the non perturbative region of the coupling constant.
It is also possible to note that the central region of the lattice,
from where the bubble started expanding, tends to assume 
energy density values close to that of the vacuum as the energy
goes away towards the borders of the lattice.
Stronger coupling constants therefore 
lead to the following quantum effect:
there is an acceleration of the energy density expansion. 
Although the two-peaked form
of the expanding waves persists, the advanced peaks (those
which arrive before to the borders of the lattice)
are strengthened with relation to the others. 
This is a clear indication that the expansion is faster 
corresponding to a non perturbative quantum effect.


What happen if the initial energy 
density excess in the center of the lattice
is increased?  In figures 8 and 9  this is shown respectively for
the classical system and for the ``complete'' (classical plus 
quantum) field with  the same set of parameters as the preceding figures
but considering $\alpha = 0.5$ in expression (\ref{IC3}).
The coupling constant 
is in the  range $\lambda = 60 fm^{-2} \simeq 236 \mu^2$.
The energy density excess is much higher than those considered before
as it can be checked by comparing with figure 8. 
In this figure the temporal evolution of the energy density 
of the condensate without quantum fluctuations is shown. 
In spite of different normalization  and energy values the 
classical dynamics is not modified with relation to figure 4 (for which
the initial condition was given by the suppression of the condensate
with expression (\ref{IC2}) ).
In figure 9 the dynamics of quantum fluctuations is taken into account.
It is also  noted in this case a quite large
 energy density amplification in the beginning of the evolution
and 
the existence of expanding regions in which the energy density
is smaller than the value of the vacuum. 
They  modify the 
two-peak structure discussed before. 
Although the last step in such 
time evolution present values for the energy density 
which may not be completely reliable due 
to numerical uncertainties the main issues are completely trustful
even because, we emphasize, the total energy is conserved. 
The acceleration found before still is present.
Moreover, it is interesting to note that there is little 
difference between the temporal evolution of cases in which 
initial conditions are given in term of $\alpha$ with different
values, i.e., for $\alpha=0.2$ or $\alpha=0.5$ in the early
time dynamics.

For the examples shown above only the short time behavior was analyzed.
These are so short time scales that the boundary conditions are not 
even relevant.
The large time behavior was briefly 
studied for some cases in the strong coupling
limit ($\lambda = 60 fm^{-2}$).
 No equilibration
 was found for larger times evolutions, i.e., the amplitude of the
 (classical and quantum) oscillations do not tend to zero. 
There are several other different approaches dealing with different 
aspects of the non-equilibrium field dynamics 
\cite{AAAB,AARTSSMIT,BERCOX,BORSZEJA}.

In the next figures another kind of (non trivial) initial 
configuration is considered for the condensate. 
In this case $\bphi(r=0)= -\bphi_0$ and $\bphi (r=10) = \bphi_0$, where
$0$ and $10$ are the borders of the lattice. 
This configuration can be seen as a wall which separates
regions with different vacua where $\bphi = \pm \bphi_0$.
By considering anti-periodic boundary
conditions we have chosen a deviation with relation to the 
KINK solution (expression \ref{IC5}) \cite{MONTMUNST}:
\be  \ba{ll}
\displaystyle{ \bphi (x,t=0) = \bphi_0 tanh \left( \frac{x-L/2}{B}
\right).
}
\ea
\ee
Where $B$ was considered to be $B \simeq 1/(4 \mu)$ and 
$\lambda = 60 fm^{-2}$.
Firstly the evolution of the equations of movement for the condensate
without quantum fluctuations ($G=0$) is performed.
In figure 10a the resulting 
classical field profile ($\bphi(t)$) at the points $x=5fm$ and $x=10fm$
are shown.
The change of the field at $\bphi(x=10,t)$  (there is a ``flip'')
is  due to the anti-periodic boundary conditions and happens exactly
at the time when the energy density bumps arrive at the border of 
the lattice. 
The field in the central region has a static and constant value given
at $t=0$.
In figure 10b we show the energy density expansion 
(it does
not have the two-peak structure due to the initial condition).
By switching on quantum fluctuations we obtain figures 11a, b, c.
In figure 11a the condensate in the same points of the lattice
as in figure 10a is shown.
The dynamics is similar but the field (at r=10 fm) 
``flips'' to the other vacuum value faster. 
In figure 11b the deviation of 
the quantum fluctuations with relation to the value in the 
vacuum, at the same points ($\delta G(t,r) = G(t,r) -G_0$), are shown. 
They clearly exhibit the energy transfer
between classical and quantum degrees of freedom. 
In particular,
when the ``condensate flip'' occurs, the quantum fluctuations are 
enhanced in the corresponding point ($r=10fm$).
For the energy density configurations, in figure 11c, there is no
strong effect with relation to the purely classical dynamics 
(seen in figure 10b).

For the figures 12 and 13 a different size of the bubble 
of suppressed condensate is considered
with the initial condition given by expression (\ref{IC2}). 
We have considered half value, i.e., $A=0.25 fm$. This yields
a smaller region out of equilibrium. 
In figure 12 the temporal evolution of the energy density of the classical 
condensate case is shown as well as in figure 13 the same
 for the classical plus quantum system.
The concentration of energy density in the ``advanced''
peaks discussed before (figures 5,7 and \cite{BRANAV}) is 
well visible in figure 13. This indicates, again, an acceleralation
of the expansion with relation to the classical dynamics.
Moreover, there is the appearance
of new bumps in the expansion due to the inclusion of the quantum dynamics
in a finite (discrete) system.


In the continuum limit the GAP equations  and the equations of movement
present the ultraviolet (UV) divergences discussed in section 2.
This is reproduced in the lattice calculations and to show 
that our results do reproduce features of the continuum 
renormalized model we have studied the limit of smaller 
lattice spacings down to $\Delta x = 0.02 fm$.
In order to continue in the scaling limit the mass $\mu^2$ must 
be changed in the same way as the GAP equations in the lattice
since it absorbs the UV infinities in the renormalization.
The two point function scales as \cite{ZINNJUSTIN}:
\be 
G^{-1} \to \frac{2 \kappa}{(\Delta x)^2 Z_R} 
\left( m^2_R + p^2 + o(p^4) \right),
\ee
where $Z_R$ is the field renormalization factor (it is finite
in the 1+1 dimensional case).
Results for the small lattice spacing limit 
are not visibly modified.
The difference is found in the normalizations 
(absolute values) of the 
two point function and of the energy density. 
By subtracting these values by the vacuum ones the
results remain unchanged. 
Besides that, the dynamics is not affected by 
these divergences, i.e., by the regularization method. 
This was shown in section 2.3 for 
the particular case of initial conditions given by small
deviations from the vacuum.

\section{SUMMARY}

We have analyzed the temporal evolution of 
expanding non-homogeneous configurations
of the \lap model considering two different approaches: the classical
equations of motion and compared its results to the equations of 
motion in the frame of the Gaussian variational approximation 
 in a 1+1 dimensional lattice.
A schematic model for the model at finite fermionic density 
has been drawn for the initial conditions and 
 the equations of movement have been solved.
The condensate may either disappear (symmetry restoration) or become
higher (no symmetric phase, with further symmetry 
breaking) at higher densities.
The parameters of the \lap model were fixed in order 
to allow a comparison between the classical and quantum 
field temporal evolution. 
We have been able to study the influence of the quantum 
fluctuations on the classical field dynamics for different sets 
of free parameters. 
By varying the parameters of 
the model and the non-homogeneous
initial conditions we have carefully investigated the expansion of 
different field configurations in the frame of the Gaussian approximation.
The quantum fluctuations
accelerate the expansion of a concentrated configuration
 of the field. 
This effect is considerable for strong coupling constants and 
particular cases of the initial conditions, namely when there
is an enhancement of the condensate (stronger symmetry breaking)
 instead of suppression (restoration). 
However, no big differences were found for these
two different initial conditions in the early time dynamics.
Closely related works  have been done by \cite{BOVEHO,SSV,BEPASA} 
respectively for other initial conditions, as for instance 
a Gaussian configuration, and
additional averaging over ensembles of mean fields
(which seems to lead to thermalization at long times). 
It is possible to conclude that the initial conditions
play an important role in the temporal evolution.
We have been concerned mainly with short time intervals evolution
and other issues related to thermalization and equilibration have
not been addressed extensively. 

{\bf Acknowledgements}\\
This work was supported by FAPESP- Brazil. 
F.L.B. wishes to thank F.S. Navarra
for several interesting discussions and a collaboration.
The numerical calculations were performed in the machines of
the Laboratory for Computation of the University of S\~ao Paulo - LCCA-USP.

\centerline{\large \bf Figure Captions}

\begin{itemize}

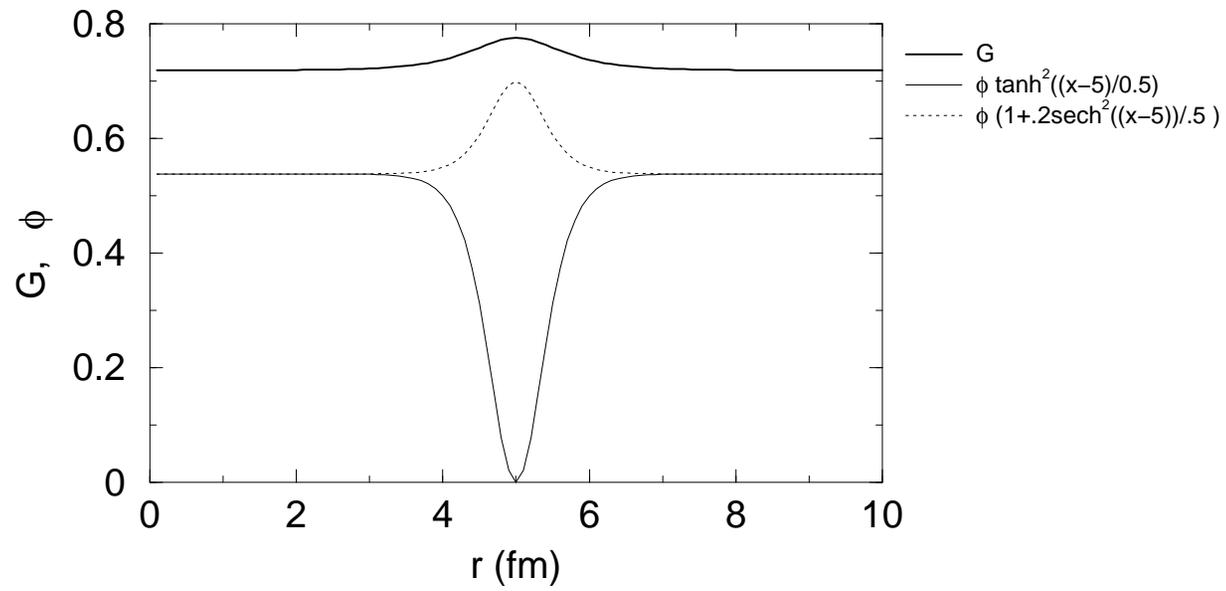
\item[{\bf Fig. 1}] Examples of initial field configuration. 
Thick solid line corresponds to the two point function $G(r,t=0)$ for 
the initial condition (\ref{IC1}) considering $\mu=150 MeV$. Thin solid and
dotted lines correspond to the initial condensate configuration of 
expressions
(\ref{IC2}) and  (\ref{IC3} with $\alpha=0.2$) respectively.

\item[{\bf Fig. 2}] Evolution of the energy density of 
the field (classical + quantum) at different times.
The initial time mass  
configuration is given by (\ref{IC1}) and $\lambda = \mu^2/12$.

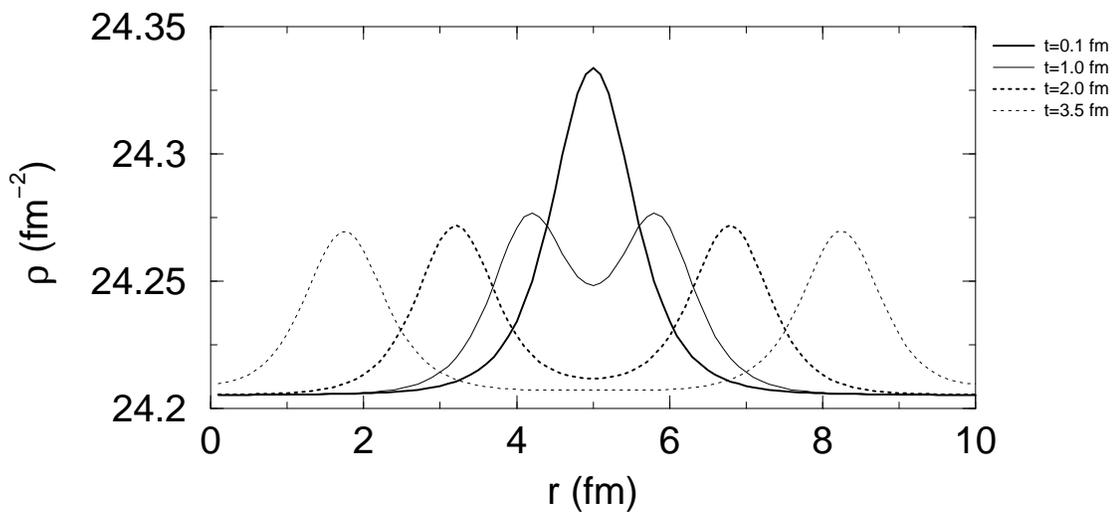
\item[{\bf Fig. 3}] The same as Fig.2 with $\lambda = 600 fm^{-2}$.

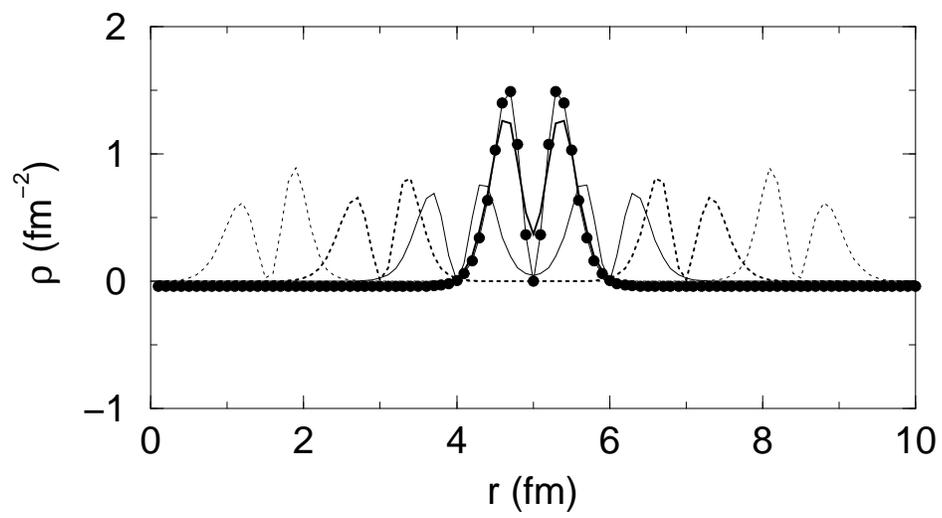
\item[{\bf Fig. 4}] Evolution of the energy distribution of 
the classical field for the initial configuration 
(\ref{IC2}) and $\lambda = 0.6 fm^{-2}$.

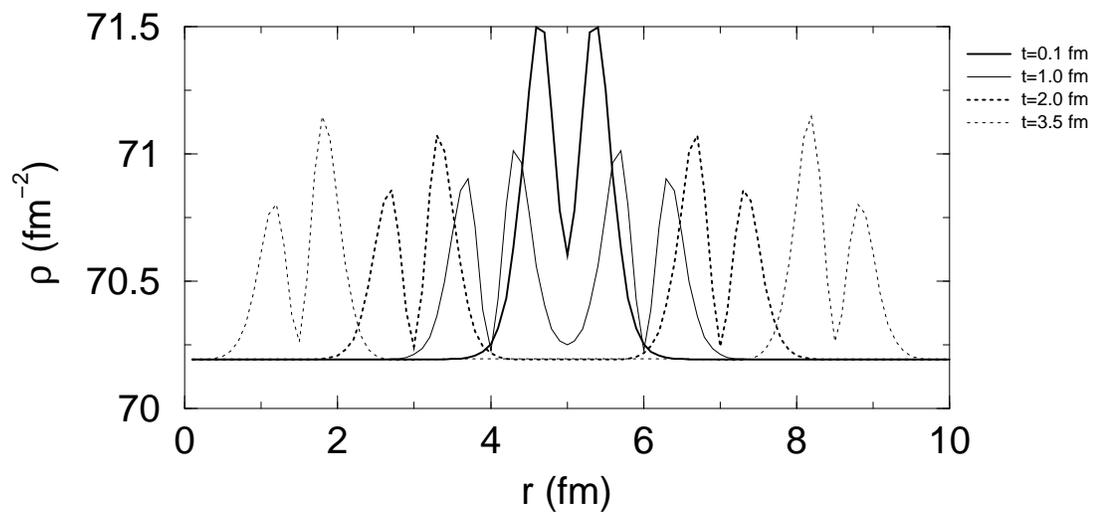
\item[{\bf Fig. 5}] Evolution of the energy distribution of 
the field with  the quantum fluctuations for 
initial configuration (\ref{IC2}) and $\lambda = 0.6 fm^{-2}$.

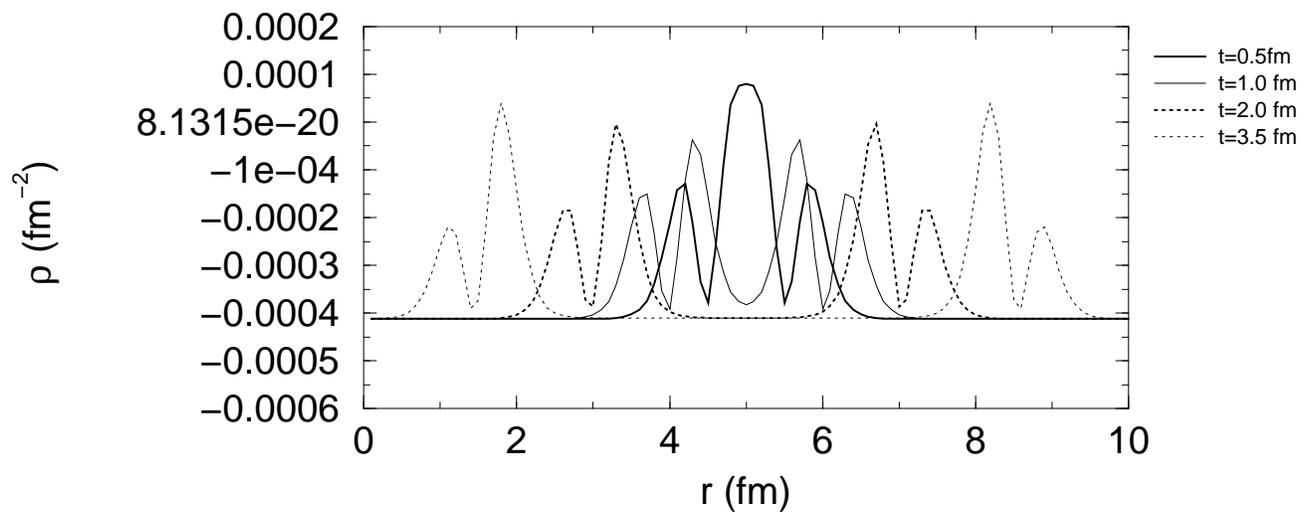
\item[{\bf Fig. 6}] Evolution of the energy distribution of 
the classical field for the initial configuration
(\ref{IC3}) with $\alpha=0.2$ and 
for  $\lambda = 60 fm^{-2}$.

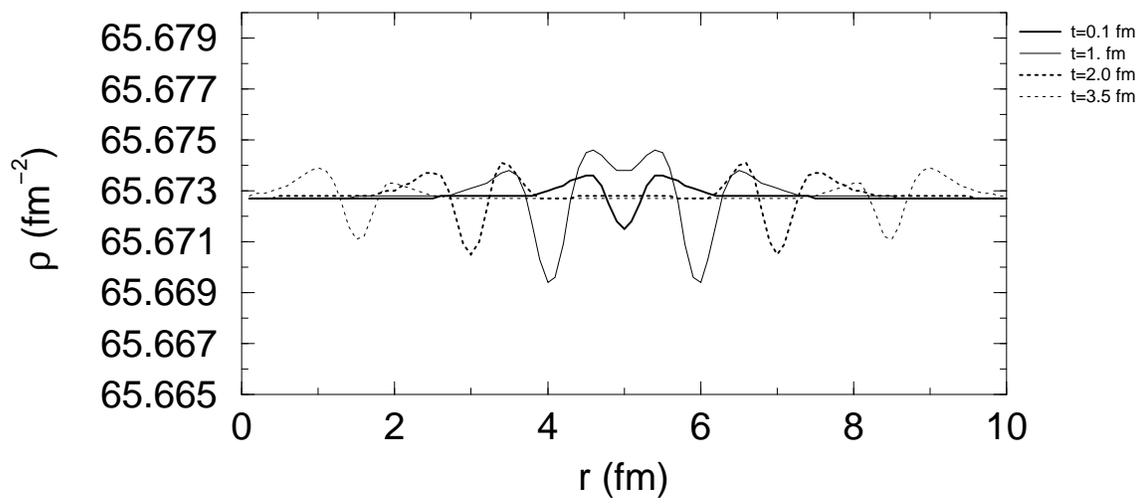
\item[{\bf Fig. 7}] Evolution of the energy distribution of 
the  field with  the quantum fluctuations for 
initial configuration (\ref{IC3}) with $\alpha = 0.2$ 
 and $\lambda = 60 fm^{-2}$.

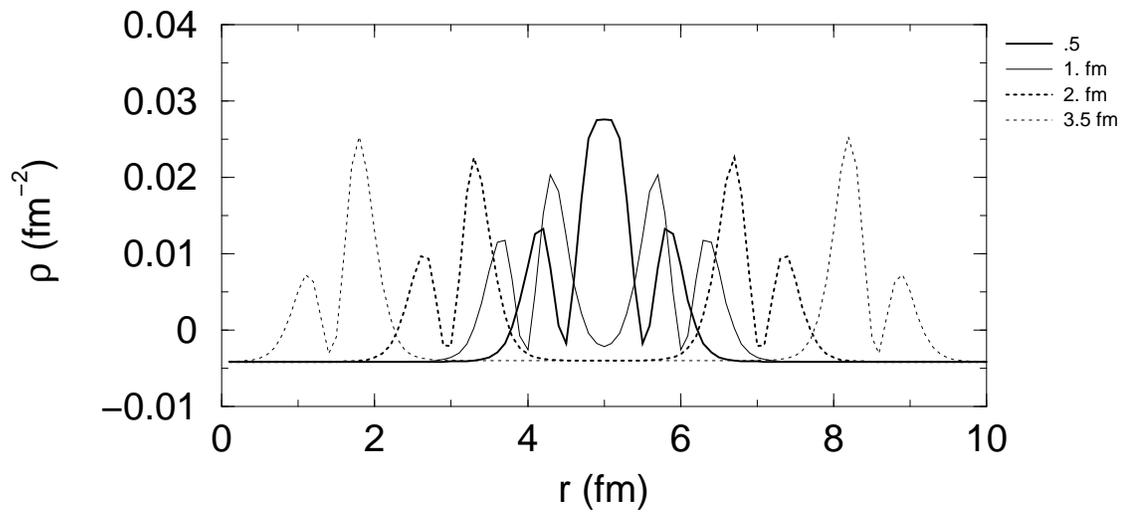
\item[{\bf Fig. 8}] Evolution of the energy distribution of 
the classical field for the initial configuration (\ref{IC3})
 $\alpha=0.5$.

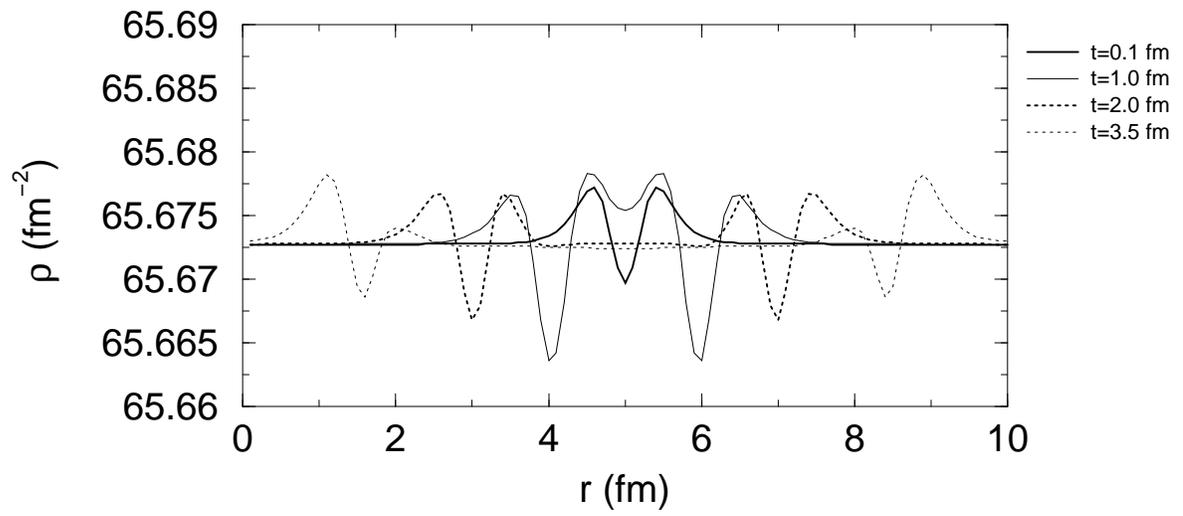
\item[{\bf Fig. 9}] Evolution of the energy distribution of 
the  field with  the quantum fluctuations for 
initial configuration (\ref{IC3}) with $\alpha = 0.5$ 
 and $\lambda = 60 fm^{-2}$.

\item[{\bf Fig. 10}] Evolution of the energy distribution of 
the classical field $\bphi(t,x)$  for the initial 
configuration (\ref{IC4}) and $\lambda=60 fm^{-2}$.

\item[{\bf Fig. 11}] Evolution of the energy distribution of 
the  field with  the quantum fluctuations for 
initial configuration (\ref{IC4}) with 
 $\lambda = 60 fm^{-2}$.

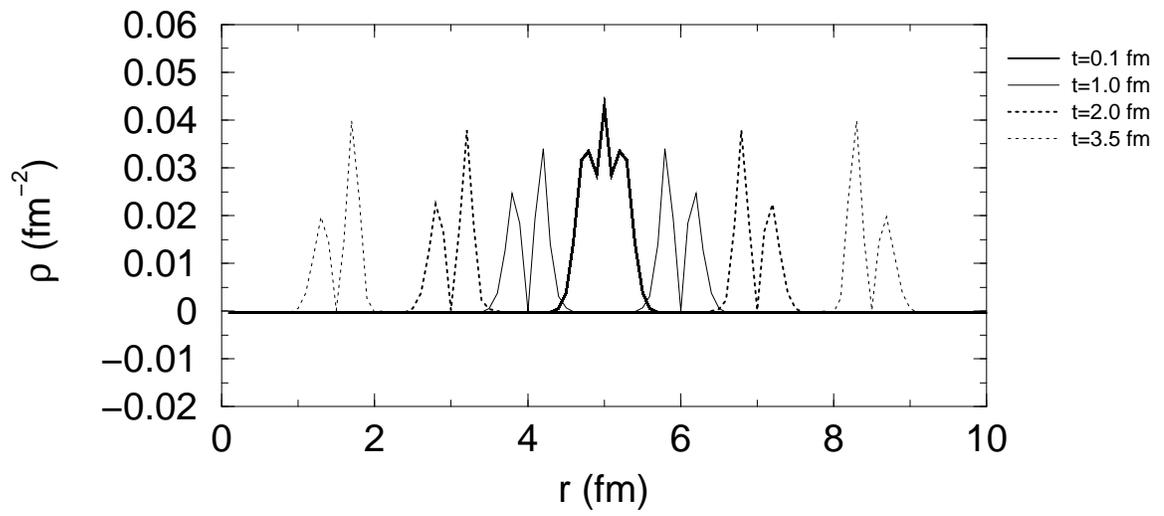
\item[{\bf Fig. 12}] Evolution of the energy distribution of 
the classical field for the initial configuration 
(\ref{IC2})  with $A=0.25 fm$.

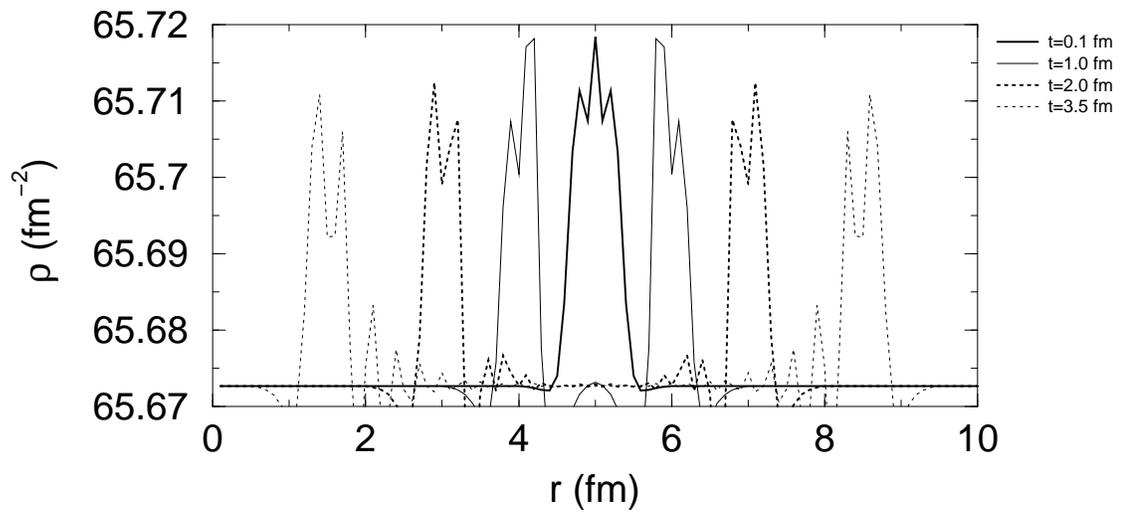
\item[{\bf Fig. 13}]  Evolution of the energy distribution of 
the field with the quantum fluctuations for 
initial configuration (\ref{IC2}) with $A=0.25 fm$.

\end{itemize}


\begin{thebibliography}{11}
\bibitem{HYDRO} U. Heinz, hep-ph/9902424; 
       F.Grassi, Y. Hama and T. Kodama, Z. Phys. {\bf C 73} 153, (1996);  
                      and references therein.
\bibitem{COSMO} D.H. Lyth, A. Riotto, Phys. Rept.{\bf 314}, 1 (1999).
\bibitem{CONDMATT} A.J. Heeger,S. Kivelson, J.R. Schrieffer, W.P. Su, 
                 Rev. Mod. Phys. {\bf 60}, 781 (1988); 
                  G. Gr\"uner, Rev. Mod. Phys. {\bf 60}, 1129 (1988).

\bibitem{EBOJAPI} O. \'Eboli, R. Jackiw and So Young Pi, Phys. Rev. 
                     {\bf  D37}, 3557 (1988).
\bibitem{COPISTA} F. Cooper, S.-Y. Pi and P.N. Stancioff, Phys. Rev.
                    {\bf D 34}, 3831 (1986).

\bibitem{COHAKLUMO}  F. Cooper, S. Habib, Y. Kluger, E. Mottola, Phys.
                     Rev. {\bf D 55}, 6471 (1997).       

\bibitem{BOVE1} D. Boyanovsky, M.D'Attanasio, H. J. de Vega, R. Holman,
                  D.-S. Lee, Phys. Rev. {\bf D 52}, 6805 (1996);
D.Boyanovsky, H. deVega, R. Holman, D.S. Lee, A. Singh, Phys. Rev.
 {\bf D 51}, 4419 (1995).

\bibitem{BOVEHO} D. Boyanovsky, H.J.  
        de Vega, R. Holman, Phys. Rev. {\bf D 51},
                 734 (1995); Phys. Rev. {\bf D 54}, 1748 (1996);
                D. Boyanovsky, F. Cooper, H.J. de Vega, P. Sodano, 
       hep-ph/9802277.

\bibitem{FLB98a} F.L. Braghin, Phys. Rev.{\bf D 57}, 3548, (1998).

\bibitem{BAHEPA} J. Baacke, K. Heitmann, C. P\"atzold, Phys. Rev. 
               {\bf D 57}, 6406 (1998);
                J. Baacke, K. Heitmann, C. P\"atzold, 
                    Phys. Rev. {\bf D 57} 6406, (1999); 
                    J. Baacke and K. Heitmann, hep-ph/0003317.

\bibitem{SSV} M. Sall\'e, Jan Smit and J. C. Vink, 
                hep-ph/0012362; hep-ph/0012346.

\bibitem{BEGLERAM} A. Berera, M. Gleiser, R.O. Ramos, Phys. Rev. {\bf D 58},
                     123508-1 (1998).
\bibitem{AAAB} G. Aarts, G.F. Bonini, C. Wetterich, hep-ph/0003262;
               hep-ph/0007357; G. Aarts and J.Berges, hep/0103049;
                J. Berges, hep-ph/0105311.

\bibitem{TSUE} Y. Tsue, A. Koike, N. Ikezi, hep-ph/0103246.
\bibitem{BRANAV} F.L. Braghin and F.S. Navarra, 
                Phys. Lett. {\bf B 508}, 243 (2001).
\bibitem{FLB2001b} F.L. Braghin, {\it in preparation}.
\bibitem{DUMITRU}  A. Dumitru et al., nucl-th/0010107.

\bibitem{AARTSSMIT} G. Aarts and J. Smit, hep-ph/9906538.

\bibitem{COHAKLUMOPAAND}  F. Cooper, S. Habib, Y. Kluger, E. Mottola, 
                          J.P. Paz and P.R. Anderson, Phys.
                          Rev. {\bf D50}, 2848 (1994).
\bibitem{SAPI} Pi,S.Y. , M. Samiullah, Phys. Rev. {\bf D 36}, 3128 (1987).

\bibitem{BERCOX} J. Berges, J. Cox, hep-ph/0006160.

\bibitem{JACKER} R. Jackiw and A. K. Kerman, Phys. Lett {\bf A71}, 158 (1979).
\bibitem{COOPMOT} F. Cooper and E. Mottola, Phys. Rev. {\bf D36}, 3114 (1987).

\bibitem{BARGAN}  T. Barnes and G.I. Ghandour, Phys. Rev. {\bf D22}, 924
                  (1980).

\bibitem{PIZA} P.L. Natti, A.F.R. de Toledo Piza, Phys. Rev. {\bf D 54},
                 7867 (1996).

\bibitem{AARTS} G. Aarts, G.F. Bonini and C. Wetterich, hep-ph/0007357.

\bibitem{BEPASA} L.M.A. Bettencourt, K. Pao, J.G. Sanderson,
                 hep-ph/0104210
\bibitem{DEMA} C. Destri, E. Manfredini, hep-ph/0001177; hep-ph/0001178.
\bibitem{RSSB} L. Dolan and R. Jackiw, Phys. Rev. {\bf D 9}, 3321 (1974);
               G. Bimonte, D. Iniguez, A. Tarancon,
                 C.L. Ullod, Nucl. Phys. {\bf B 515}, 345 (1998).

\bibitem{BAABOVE} J. Baacke, D. Boyanovsky, H. de Vega, hep-ph/9907337.

\bibitem{BLARIP}  J.P. Blaizot and G. Ripka, {\it Quantum Theory of 
Finite Systems}, The MIT Press, Cambridge, (1986).

\bibitem{NRSSB} S. Weinberg, Phys. Rev. {\bf D 9}, 3357 (1974);
                M.B. Pinto, R.O. Ramos, Phys. Rev. {\bf D 61}, 125016 (2000).

\bibitem{ZINNJUSTIN} J. Zinn Justin, {\it Quantum Field Theory and 
                    Critical Phenomena}, 3rd edition, Oxford University
                   Press, Oxford (1996). 

\bibitem{STEVENSON} P.M. Stevenson, Phys. Rev. {\bf D 32}, 1389 (1985).
                     A.K. Kerman, C. Martin, D. Vautherin, Phys. Rev. 
                       {\bf D 47}, 632 (1993).

\bibitem{MONTMUNST} I. Montvay and G. M\"unster, {\it Quantum Fields
                   on a Lattice}, Cambridge Univ. Press, Cambridge (1984).

\bibitem{BORSZEJA}  Sz. Bors\'anyi, Zs. Sz\'ep, hep-ph/0011283;
 A. Jakov\'ac, A. Patk\'os, P. Petreczky, Zs. Sz\'ep,
Phys. Rev. {\bf D 61}: 025006 (2000).
                 

\end{thebibliography}
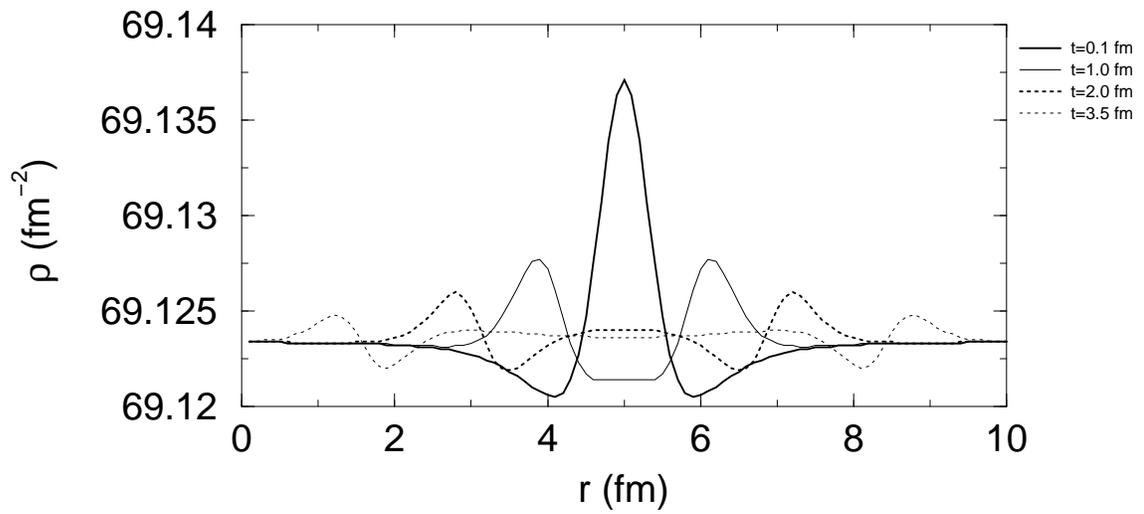
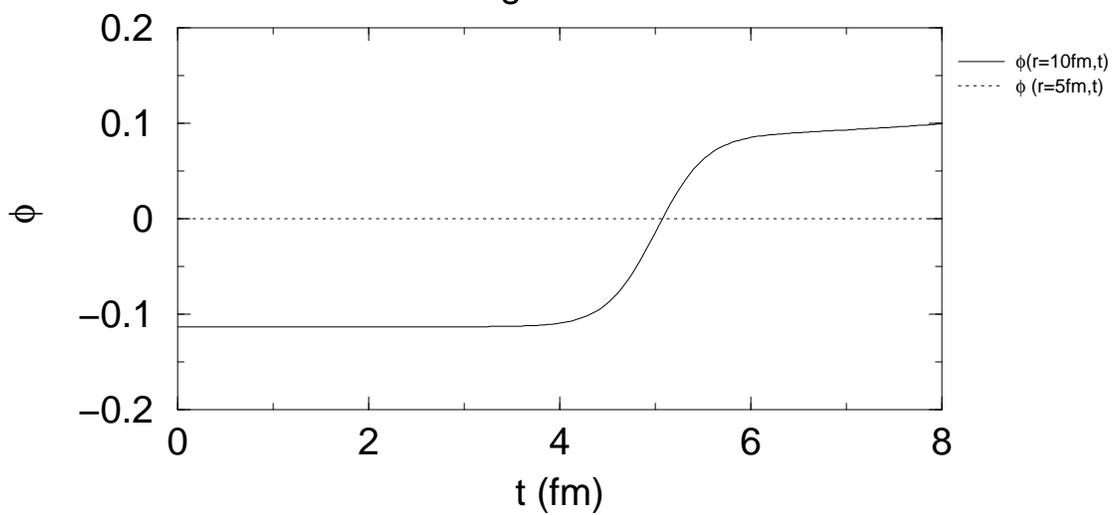
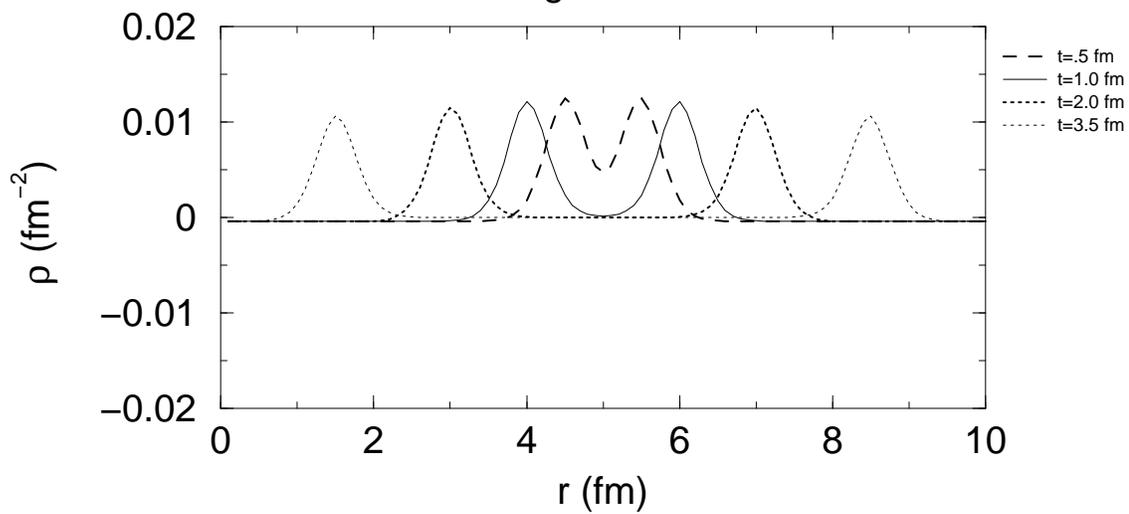
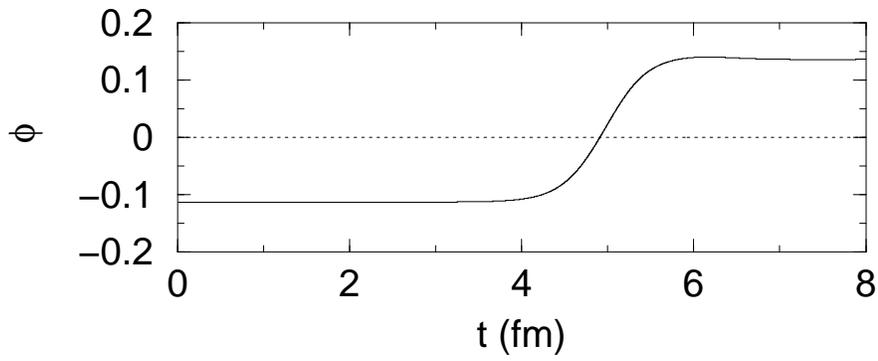
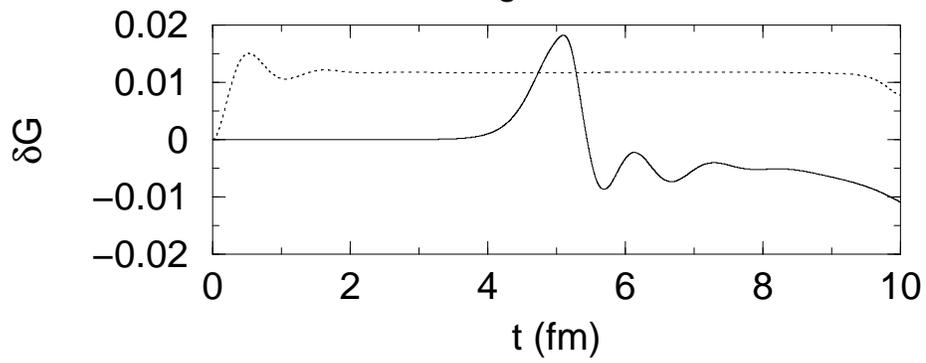
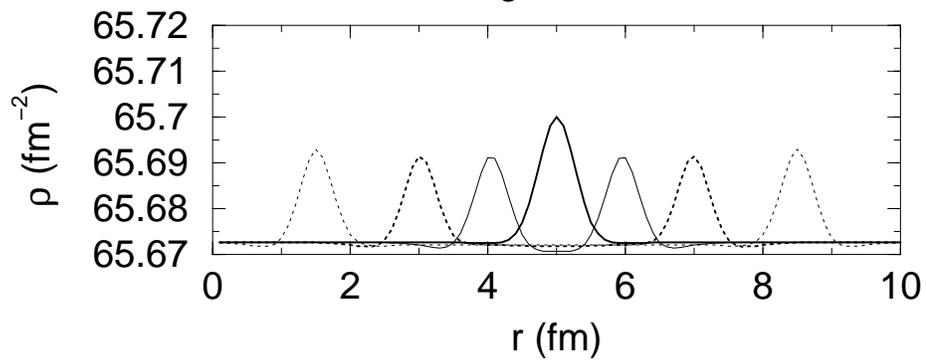
\end{document}